\documentclass [reprint,aps,amsmath,amssymb,twoside,prl,showkeys,superscriptaddress]{revtex4-1}
\usepackage{verbatim}
\usepackage{graphicx}
\usepackage[T1]{fontenc}
\usepackage{textcomp}
\usepackage{txfonts}
\usepackage{tensor}
\usepackage{cancel}
\usepackage{perpage} %the perpage package
\MakePerPage{footnote} %the perpage package comman
\usepackage{color}
\newcommand{\HCd}{\mathcal{H}}

\def\HCdt0{\tilde{\HCd}_{0}}
\newcommand{\afffias}{Frankfurt Institute for Advanced Studies (FIAS), Ruth-Moufang-Strasse~1, 60438 Frankfurt am Main, Germany}

\newcommand{\affbgu}{Physics Department, Ben-Gurion University of the Negev, Beer-Sheva 84105, Israel}
\newcommand{\affbahamas}{Bahamas Advanced Study Institute and Conferences, 4A Ocean Heights, Hill View Circle, Stella Maris, Long Island, The Bahamas}

\begin{document}
\title{Two scalar fields inflation from scale-invariant gravity with modified measure}
\author{David Benisty}
\email{benidav@post.bgu.ac.il}
\affiliation{\afffias}\affiliation{\affbgu}
\author{Eduardo I. Guendelman}
\email{guendel@bgu.ac.il}
\affiliation{\afffias}\affiliation{\affbgu}\affiliation{\affbahamas}
\date{\today}
\begin{abstract}
A new class of gravity-matter model defined with an independent non-Riemannian volume form is studied, in the second order formalism. The action has a global scale invariance symmetry, which can be broken by the equation of motion of the measure fields. By a conformal transformation the theory transforms into a theory which governed by two independent scalar fields and a specific potential. When the scale invariance symmetry is not broken also in the equations of motion, only one field appear in the effective potential. This effective potential has a flat region which is responsible for the inflation, and it ends with a minimum, which is responsible for the present vacuum energy. In the general case, with a spontaneous symmetry breaking of the scale symmetry, one scalar field is responsible for the inflation and the other could be responsible for the particle creation. The first field (inflaton) can slowly roll, while the second field (curvaton) is more strongly coupled to the potential. The number of e-folds for both models is also dissuaded and could be constraint in the future.
\end{abstract}
\maketitle
\section{Introduction}
Since the 1980s, developments in cosmology have been influenced to a great extent by the idea of inflation \cite{Starobinsky:1979ty}-\cite{Albrecht:1982wi}, which provides an attractive scenario for the solution of the fundamental puzzles of the standard Big Bang model, like the horizon and the flatness problems as well as providing a framework for sensible calculations of primordial density perturbations \cite{Mukhanov:1981xt}-\cite{Guth:1982ec} Especially the pioneer Starobinsky model ($R+ \alpha R^2$) \cite{Starobinsky:1980te} which still remains viable and produces one from the best bit to existing observational data compared to other inflationary models \cite{Akrami:2018odb}. However, although the inflationary scenario is very attractive, it has been recognized that a successful implementation requires some very special restrictions on the dynamics that drive inflation. In particular, in New Inflation, a potential with a large flat region, which then drops to zero (or almost zero) in order to reproduce the vacuum with almost zero (in Planck units) cosmological constant of the present universe, is required. It is hard to find a theory that gives a potential of this type naturally.  

The concept of scale invariance appears as an attractive possibility for a fundamental symmetry of nature. Dimensionless coupling constants for example appear related to good renormalizability properties. In its most naive realizations, such a symmetry is not a viable symmetry however, since nature seems to have chosen some typical scales. Here we will find that scale invariance can nevertheless be incorporated into realistic, generally covariant field theories. However, scale invariance has to be discussed in a more general framework than that of standard generally relativistic theories, where we must allow in the action, in addition to the ordinary measure of integration $\sqrt{-g}$, another measure $\Phi$, which is a density built out of degrees of freedom
independent of that of the metric $g_{\mu\nu}$.

To achieve global scale invariance, also a "dilaton" $\phi$ has to be introduced. As was discussed in \cite{Guendelman:1999qt}, a potential consistent with scale invariance can appear for the $\phi$ field. Such a potential has a shape which makes it suitable for the satisfactory realization of an inflationary scenario. Alternatively, it can be of use in a slowly rolling $\Lambda$ scenario for the late universe. The equations of motion of the fields that define the measure give us the possibility of a spontaneous breaking of the scale invariance. In \cite{Guendelman:1999qt} these issues were studied in the first order formalism, where the new measure $\Phi$ ends up not being a new degree of freedom, but is instead determined by a constraint equation. In contrast, here we will use the second order formulation, where the connection is defined as the Christoffel symbol, and as a consequence, the field $\chi= \frac{\Phi}{\sqrt{-g}}$ provide us another scalar, which could play a role of a "curvaton field" \cite{Enqvist:2001zp}-\cite{Moroi:2001ct}, while the dilaton $\phi$ can play the role of inflaton (as was studied in the previous case studied in the first order formalism). In cosmological scenarios based on modified measure theory, in the first order formalism, the curvatun has to be added as an additional independent field \cite{Guendelman:2015liz}. As we will see, in the second order formalism this additional fields is generated naturally from the modified measure itself. 

Multi-field inflation with a curved scalar geometry has been found to support background trajectories that violate the slow-roll, slow-turn conditions and thus have the potential
to evade the  constraints put forward by the proponents of the string theory swampland, a proposal to differentiate low energies theories that arise from string theory from those which do not \cite{Vafa}, as was done in  \cite{Achucarro:2018vey}. In \cite{Achucarro:2018vey} it was shown that in multifield scalar cosmologies, one can still have something that resembles slow roll, while being consistent with swamp constraint, presumably that gets us back to an almost constant vacuum energy for the present universe (as well as in the inflationary phase) while the still staying in the landscape instead of "falling into the swamp", if one wishes to do so. 

In addition, in two field inflation,  one field could be used as an inflaton field and the other as a curvaton. In recent years we have seen the introduction of various multi-field inflationary scenarios. For instance, \cite{Christodoulidis:2019mkj} unifies the late-time attractors of hyperinflation, angular, orbital and side-tracked inflation. \cite{Christodoulidis:2019jsx} discovers that dynamical
bifurcations play an integral role in the transition between geodesic and non-geodesic motion and discuss the ability of scaling solutions to describe realistic multi-field models. \cite{Vicentini:2019etr} 
studies a scale invariance which spontaneously broken and a mass scale naturally emerges and one recovers to the usual Einstein-Hilbert action.

Earlier publications have shown that in the metric-affine formalism, in the presence of $R^2$ term for scale invariant action with the modified measure, produces inflation with two flat regions  for the potential: one for the inflation era and one for the dark energy dominant era \cite{Guendelman:2002js}-\cite{delCampo:2010kf}. In addition breaking of scaling invariance symmetry in the standard model combined with spontaneous symmetry breaking of gauge symmetry was studied in \cite{Guendelman:2018xcu}-\cite{Guendelman:2017axj}. These are examples to theories which uses the spontaneous symmetry breaking from the modified measure and will be discussed here. In addition, one from fundamental features that this model could achieve is the quintessential inflation: which the same fields has the property of inflation for the early universe and approach dark energy in the late universe \cite{Zhang:2019ing}.

\section{Modified measure theories}
Many modified theories of gravity have been formulated for explaining phenomena beyond General Relativity. One example is the two measures theory \cite{TMT4}-\cite{TMT9} where in addition to the regular measure of integration in the action $ \sqrt{-g} $ includes another measure of interaction which is also a density volume and a total derivative. In this case, one can use for constructing this measure 4 scalar fields $ \varphi_{a} $, where $ a=1,2,3,4 $. Then, we can define the density:
\begin{equation}
\Phi=\varepsilon^{\alpha\beta\gamma\delta}\varepsilon_{abcd}\partial_{\alpha}\varphi_{a}\partial_{\beta}\varphi_{b}\partial_{\gamma}\varphi_{c}\partial_{\delta}\varphi_{d} 
\end{equation}
and then we can write an action that uses both of these densities:
\begin{equation}\label{2}
        S=\int d^{4}x \,\Phi\mathcal{L}_{1}+\int d^{4}x\,\sqrt{-g}\mathcal{L}_{2}\,.
\end{equation}
As a consequence of the variation with respect to the scalar fields $ \varphi_{a} $, assuming that $ \mathcal{L}_{1} $
and $ \mathcal{L}_{2} $
are independent of the scalar fields $\varphi_{a}$, we obtain that
\begin{equation} \label{measure}
        A_{a}^{\alpha}\partial_{\alpha}\mathcal{L}_{1}=0\,,
\end{equation}
where $ A_{a}^{\alpha}=\varepsilon^{\alpha\beta\gamma\delta}\varepsilon_{abcd}\partial_{\beta}\varphi_{b}\partial_{\gamma}\varphi_{c}\partial_{\delta}\varphi_{d} $. As a consequence of the variation with respect to the scalar fields $ \varphi_{a} $, assuming that $ \mathcal{L}_{1} $ and $ \mathcal{L}_{2} $ are independent of the scalar fields $\varphi_{a}$, we obtain that for $ \Phi\neq0 $ it implies:
\begin{equation}
\mathcal{L}_{1}=M=\textbf{const}
\end{equation}
that $\mathcal{L}_{1}$ is a consonant in the equations of motion.

The study of Two Measure Theories in the second order formalism was done in \cite{TMTso1}-\cite{TMTso2}, although it was not done in a connection to scaling invariance symmetry. In this paper we investigate the second order solution under scale invariant theory.   
\section{Scale invariant modified gravity}
Our starting point is the general action, which is contains both measures:
\begin{equation}
\begin{split}
\mathcal{L} = \int \Phi (-\frac{1}{\kappa} R +\frac{1}{2} g^{\mu\nu}\phi_{,\mu}\phi_{,\nu} - V(\phi)) \\+ \sqrt{-g} U(\phi) +  \lambda \frac{\Phi^2}{\sqrt{-g}},
\end{split}
\end{equation}
and slightly generalized the form of Eq. (\ref{2}) to allow the $\Phi^2$ term, which is still consistent with global scale invariance. The $R$ is the standard Ricci scalar curvature defined from the Levi-Civita connection, that is will used in the second order formulation.

The global scale invariance symmetry behavior for the scalar field $\varphi_a$, which is defines the measure $\Phi$, is:
\begin{equation}
\varphi_a\rightarrow \lambda_{ab} \, \varphi_b(\varphi)
\end{equation}
where $\lambda_{ab}$ is a constant matrix. From here we see that the  global scale invariance symmetry for the modified measure is:     
\begin{equation}
\Phi \rightarrow \lambda \, \Phi
\end{equation}
where $\lambda$ is the determinant:
\begin{equation}
\lambda = \det \left[\lambda_{ab}\right]
\end{equation}
Furthermore the metric and connection transform as follows:
\begin{equation}
g_{\mu\nu} \rightarrow \lambda g_{\mu\nu} ,\quad \Gamma_{\nu\lambda}^{\mu} \rightarrow \Gamma_{\nu\lambda}^{\mu}
\end{equation}
with the same $\lambda$ that defined before. In order to maintain the global scale transformation, the potentials has to be taken in the form of:
\begin{equation}
V(\phi) = f_1 \exp(-\alpha \phi), \quad U(\phi) = f_2 \exp(-2\alpha \phi)
\end{equation}
where the transformation for the dilaton field $\phi$ is:
\begin{equation}
\phi \rightarrow \phi + \frac{\ln \lambda}{\alpha}
\end{equation}
Under those symmetries, the action is scale invariant. 
  \begin{figure*}[t!]
 	\centering
 \includegraphics[width=1\textwidth]{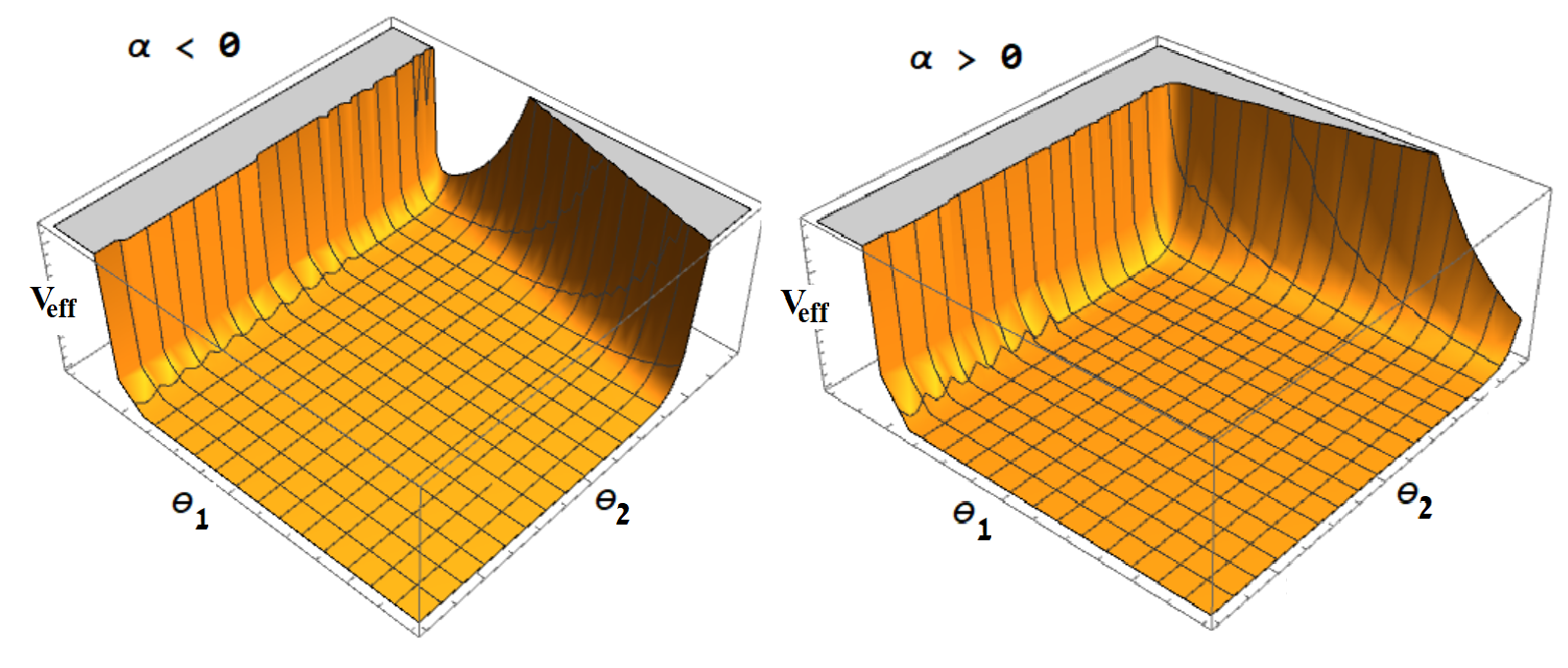}
 \caption{The effective potential $V_\textbf{eff}$ vs. the scalar fields $\theta_1$ and $\theta_2$ for $M>0$ and different signs of $\alpha$. For $\alpha < 0$ there is always a minimum in the $\theta_1$ direction regardless of the value of $\theta_1$.} 
\label{fig1}
 \end{figure*} 
Matter fields can be also introduces in a scale invariant way. In Ref. \cite{Blas:2011ac} a density independent on the metric was also introduced in a scale invariant way, but with no dilaton $\phi$ field. In \cite{Blas:2011ac} powers of the ratio between the metric independent density and the $\sqrt{-g}$ were consider as a way to generate potentials for the scalar that is derived from the ratio of between the metric independent density and the $\sqrt{-g}$. In contrast, here the origin of the scalar potential are the original dilaton potentials $U(\phi)$ and $V(\phi)$, and $\frac{\Phi^2}{\sqrt{-g}}$ which only generates a constant term in the effective potential. The $\Phi^2$ term has an interpretation of an ordinary kinetic term for the measure fields.      
\subsection{The equations of motion}
There are three independent equation of motion. The equation of motion with respect to the modified measure gives:
\begin{equation}\label{eomm}
-\frac{1}{\kappa} R +\frac{1}{2} g^{\mu\nu}\phi_{,\mu}\phi_{,\nu} - V(\phi) + 2\chi \lambda= M
\end{equation}
where $M$ is an integration constant, and $\chi$ is defined to be the fraction between the measures:
\begin{equation}
\chi = \frac{\Phi}{\sqrt{-g}}
\end{equation}
The Eq. (\ref{eomm}) spontaneously breaks the scale invariance, because for $M \neq 0$ the left hand side transforms under scale transformation, while the right hand side does not.

The second variation, with respect to the dilaton field $\phi$ gives:
\begin{equation}\label{sf}
-\phi_{,\alpha} \chi^{,\alpha} + \chi (V'(\phi)-\Box \phi) - U'(\phi) = 0
\end{equation}
which generalizes Klein-Gordon equation. The last variation, with respect to the metric, give the field equation:
\begin{equation}\label{veq}
\frac{1}{\kappa} (R_{\mu\nu}+ g_{\mu\nu} \frac{\Box \chi}{\chi} - \frac{\chi_{,\mu;\nu}}{\chi}) = \frac{1}{2}\left(\phi_{,\mu}\phi_{,\nu} - g_{\mu\nu} \frac{U(\phi)}{\chi}\right)
\end{equation}
In order to construct the Einstein Tensor $G_{\mu\nu}$, we combine the Ricci tensor and Ricci scalar from Eq. (\ref{eomm}) and Eq. (\ref{veq}): 
\begin{equation}\label{EinTen}
\begin{split}
\frac{2}{\kappa}\left(G_{\mu\nu} + g_{\mu\nu} \frac{\Box \chi}{\chi} - \frac{\chi_{,\mu;\nu}}{\chi}\right) \\ = T_{\mu\nu}^{(\phi)} + g_{\mu\nu} \left( V(\phi) - \frac{U(\phi)}{\chi} + M \right) + g_{\mu\nu} \chi^2 \lambda 
\end{split}
\end{equation}
where $T_{\mu\nu}^{(\phi)}$ is the stress energy momentum tensor of the dilaton field:
\begin{equation}
T_{\mu\nu}^{(\phi)} =\phi_{,\mu}\phi_{,\nu} - \frac{1}{2} g_{\mu\nu}g^{\alpha\beta} \phi_{,\alpha}\phi_{,\beta}
\end{equation} 
\begin{comment}
The trace equation:
\begin{equation}
-\frac{1}{\kappa}R = \frac{3}{\kappa} \frac{\Box \chi}{\chi} - \frac{1}{2} \phi_{,\mu}\phi^{,\mu}+ 2  \left(V(\phi) - \frac{U(\phi)}{\chi} + M \right) + 4\chi^2\lambda
\end{equation}
Inserting Eq. (\ref{eomm}): 
\begin{equation}
0= \frac{3}{\kappa} \frac{\Box \chi}{\chi} + V(\phi) - 2\frac{U(\phi)}{\chi}+ M
\end{equation}
\end{comment}
\subsection{Einstein frame and the two scalar field potential}
To simpling the Einstein tensor, lets use the conformal transformation \cite{Dabrowski:2008kx} for the metric, which includes the field $\chi$:
\begin{equation}
g_{\mu\nu} = \chi^{-1} \tilde{g}_{\mu\nu}
\end{equation}
This transformation cause Eq. (\ref{sf}) to take the form of:
\begin{equation}
\chi V'(\phi) = \chi^2 \tilde{\Box} \phi + U'(\phi)
\end{equation}
where $\tilde{\Box} = \frac{1}{\sqrt{-\tilde{g}}}\partial_{\mu} (\sqrt{-\tilde{g}}\tilde{g}^{\mu\nu}\partial_{\nu})$. The transformation give the basic identities:
\begin{equation}
\begin{split}
G_{\mu\nu} = \tilde{G}_{\mu\nu} - \frac{1}{2\chi^2}(\chi_{,\mu}\chi_{,\nu}-\frac{5}{2} \tilde{g}_{\mu\nu}(\tilde{g}^{\alpha\beta}\chi_{,\alpha}\chi_{,\beta})) \\
+ \frac{\chi_{\tilde{,}\mu;\nu}}{\chi} - \tilde{g}_{\mu\nu}\frac{\tilde{\Box}\chi}{\chi}
\end{split}
\end{equation}
\begin{equation}
\frac{\chi_{,\mu;\nu}}{\chi} = \frac{\chi_{\tilde{,}\mu;\nu}}{\chi} + \frac{1}{\chi^2}\tilde{T}_{\mu\nu}^{(\chi)}
\end{equation}
\begin{equation}
g_{\mu\nu}\frac{\Box \chi}{\chi} = \tilde{g}_{\mu\nu}\frac{\tilde{\Box} \chi}{\chi} - \tilde{g}_{\mu\nu}\frac{\tilde{g}^{\alpha\beta}\chi_{,\alpha}\chi_{,\beta}}{\chi^2}
\end{equation}
where:
\begin{equation}
T_{\mu\nu}^{(\chi)}=\chi_{,\mu}\chi_{,\nu} - \frac{1}{2} g_{\mu\nu}g^{\alpha\beta} \chi_{,\alpha}\chi_{,\beta}
\end{equation} 
is the kinetic energy momentum tensor for the scalar field $\chi$. Using those identities for the effective Einstein tensor give:
\begin{equation}\label{EinTen2}
\begin{split}
\frac{2}{\kappa}\left(\tilde{G}_{\mu\nu} -  \frac{3}{\chi^2}\tilde{T}_{\mu\nu}^{(\chi)}\right) = \tilde{T}_{\mu\nu}^{(\phi)} + \tilde{g}_{\mu\nu} V_{\textbf{eff}}(\phi,\chi)
\end{split}
\end{equation}
a new stress energy momentum tensor for the scalar field $\chi$, 
with the effective potential:
\begin{equation}
V_{\textbf{eff}} = \frac{1}{\chi} \left( V(\phi) - \frac{U(\phi)}{\chi} + M \right) + \lambda
\end{equation}
By setting the scalar field $\chi$ into the scalar:
\begin{equation}
\varphi = \sqrt{3} \ln \chi
\end{equation}
the Einstein tensor get the standard form:
\begin{equation}\label{EinTen3}
\begin{split}
\frac{2}{\kappa}\tilde{G}_{\mu\nu}  =\tilde{T}_{\mu\nu}^{(\varphi)} + \tilde{T}_{\mu\nu}^{(\phi)} + \tilde{g}_{\mu\nu} V_{\textbf{eff}}(\varphi,\phi)
\end{split}
\end{equation}
where $\tilde{T}_{\mu\nu}^{(\varphi)}$ is the stress energy momentum tensor of the field $\varphi$, with the effective potential: 
\begin{equation}
\begin{split}
V_{\textbf{eff}} (\varphi, \phi) =\lambda-f_1 \exp(-\frac{\varphi}{\sqrt{3}}-\alpha \phi) \\ + f_2 \exp(-\frac{2\varphi}{\sqrt{3}}-2\alpha \phi) + M \exp(\frac{\varphi}{\sqrt{3}})
\end{split}
\end{equation}
This effective potential can be re-expressed after a rotation of the fields:
\begin{equation}\label{t1}
\theta_{1} := \frac{1}{\mathcal{N}}\left(-\frac{\varphi}{\sqrt{3}}-\alpha \phi \right),
\end{equation}
and
\begin{equation}\label{t2}
\theta_{2} := \frac{1}{\mathcal{N}}\left(-\frac{\phi}{\sqrt{3}}+\alpha \varphi \right).
\end{equation}
where $\mathcal{N} := \sqrt{1/3+\alpha^2}$ is the normalization factor to guarantee that the transformation (\ref{t1}) and (\ref{t2}) is a rotation. Notice that under scale transformation $\theta_1$ is invariant, while $\theta_2$ transforms. The final effective potential gets the form:
\begin{equation}\label{pot}
\begin{split}
V_{\textbf{eff}} (\theta_1,\theta_2) =-f_1 \exp(-\mathcal{N}\theta_{1}) \\ + f_2 \exp(-2\mathcal{N}\theta_{1}) + M \exp\left[-\frac{1}{\mathcal{N}} (\frac{\theta_1}{3}-\frac{\alpha\theta_2}{\sqrt{3}})  \right] +\lambda
\end{split}
\end{equation}
This potential could investigate into two special cases. One case is $M=0$ where the potential is only depend on one scalar field $\theta_1$. Different case is when $M \neq 0$, where for $\theta_2 \rightarrow -\infty$ the contribution of $\theta_2$ is negligible. In Fig.\ref{fig1} we can see the plot for the effective potential vs. $\theta_1$ and $\theta_2$. In this plot $f_{1,2}$ as well as $M$ are positive.

\section{No scale invariant symmetry breaking}
\subsection{The shape of the potential}
In the case of $M=0$ the scale invariance symmetry is maintained also in the equations of motion. This case can be also obtain in the region $\theta_2 \rightarrow -\infty$. The effective potential takes the form:
\begin{equation}
V_{\textbf{eff}} (\theta_1,M=0) =\lambda-f_1 \exp(-\mathcal{N}\theta_{1}) + f_2 \exp(-2\mathcal{N}\theta_{1}) 
\end{equation}
Now the effective potential is depends only on the scale invariant variable $\theta_1$. 
To find the critical point, let's derivative the potential with respect to $\theta_1$ and set it to zero: 
\begin{equation}
\frac{\partial V_{\textbf{eff}}(\theta)}{\partial\theta} = 0 \quad \Rightarrow \quad \theta_m =\frac{1}{N} \ln \left(\frac{2f_2}{f_1} \right)
\end{equation}
The extreme point $\theta_m$ exists only if $f_1$ and $f_2$ have the same sign. The value of the extremal potential point is:
\begin{equation}
V_{\textbf{eff}}(\theta_m) = \lambda - \frac{f_1^2}{4f_2} := \Lambda
\end{equation}
The second derivative of the potential at this point has the value of: $\frac{f_1^2}{4f_2}$. Therefore the potential would be stable for $f_2>0$. So we need the condition:
\begin{equation}
\lambda > \frac{f_1^2}{4f_2},
\end{equation}
which cause the potential to be positive at the extreme point $V_{\textbf{eff}}(\theta_m) > 0$. The extreme point will be a minimum only for $f_1,f_2 > 0$. 

In addition to that, the asymptotic value of the potential:
\begin{equation}
\lim_{\theta_1\to\infty} V_{\textbf{eff}}(\theta_1,M=0) = \lambda
\end{equation}
the inflationary vacuum energy. 
  \begin{figure}[t!]
 	\centering
 \includegraphics[width=0.4\textwidth]{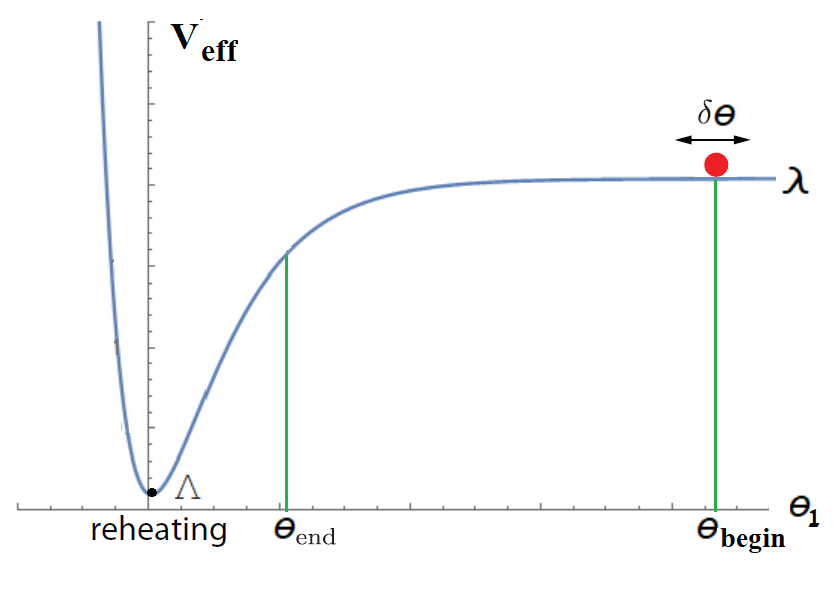}
\caption{The effective potential in the orthonormal fields under the one field approximation.} 
\label{fig2}
\end{figure}
In Fig.(\ref{fig2}) we can see that the universe begins in $\theta_\textbf{begin}$ and the value of the vacuum energy is very close to $\lambda$. The universe slow rolls in that era, until the point $\theta_\textbf{end}$ until the slow roll parameter $\epsilon \approx 1$. The final value of the vacuum energy is the present cosmological constant $\Lambda$. Let's calculate the slow roll parameters (that was defined by A. Liddle) \cite{Liddle:1992wi} for that limit and predicts the number of e-folds.  
\subsection{Slow roll inflation and the number of e-folds}
In order so calculate the number of e-Folds for that inflationary scenario, we starts with the slow roll parameter:
\begin{equation}
\epsilon \approx \frac{1}{2}\left(\frac{V_{,\theta_1}}{V}\right)^2 = \frac{1}{2} \left(\frac{\mathcal{N} e^{-\mathcal{N} \theta_1} (f_1-2f_2 e^{ -\mathcal{N}\theta_1 })}{-f_1 e^{-\mathcal{N} \theta_1 }+f_2 e^{-2 \mathcal{N}\theta_1 }+\lambda} \right)^2,
\end{equation}
and the $\eta$:
\begin{equation}
\eta \approx \frac{V_{\theta_1,\theta_1}}{V} = \mathcal{N}^2 \frac{f_1 e^{- \mathcal{N} \theta_1}-4 f_2 e^{-2  \mathcal{N} \theta_1}}{f_1 e^{- \mathcal{N} \theta_1}-f_2 e^{-2  \mathcal{N} \theta_1 }-\lambda }.
\end{equation} 
One can see that for $\theta_1 \rightarrow \infty$ the parameters are very low, if we assume that $\lambda$ has a very big value (as expected from the vacuum energy in the inflationary epoch):
\begin{equation}
\theta_1 \rightarrow \infty \quad \Rightarrow \quad \epsilon, \eta \ll 1 
\end{equation}
This shows that the slow roll assumption is valid for the area, where $\theta_{\textbf{end}}<\theta_1<\theta_{\textbf{begin}}$.

In order to estimate the end of inflation, we set the $\epsilon \approx 1$. This leads to a complicated condition. Under the assumption of $ \Lambda \ll \lambda$, which states that the present value of the vacuum energy (the cosmological constant) in much lower then the inflationary value of the vacuum energy, one can define the dimensionless quantity:
\begin{equation}
\delta := \frac{\Lambda}{\lambda} \ll 1
\end{equation}
Under this definition, the solution for $\epsilon \approx 1$ gives the solution: 
\begin{equation}
\theta_\textbf{final} = \frac{1}{\mathcal{N}}\ln\left[\frac{f_2}{f_1}\left(\sqrt{8} \mathcal{N}+2\right) \right] -\delta\frac{\sqrt{2} \mathcal{N}+1}{\sqrt{2}\mathcal{N}^2}
\end{equation}
After this point the inflation ends and the scalar fields undergoes consolations, which produce reheating of the universe. 
The number of e-folds is calculated from the equation:
\begin{equation}
\begin{split}
\textbf{NumE-folds} = \int_{\theta_\textbf{end}}^{\theta_{\textbf{Begin}}} \frac{d\theta_1}{\sqrt{2\epsilon}}
\end{split}
\end{equation}
For simplifying the solution, under the assumption of $\delta \ll 1$, the number of E-folds term is:
\begin{equation}
\begin{split}
\textbf{NumE-folds} \approx \frac{f_1 e^{\mathcal{N} \theta_{\textbf{begin}}}}{4 f_2\mathcal{N}^2}-\frac{\theta_{\textbf{begin}}}{2\mathcal{N}}\\ -\frac{f_1}{4 f_2 \mathcal{N}^2}e^{\frac{\sqrt{2} f_2 \mathcal{N}}{f_1}\left(2 \mathcal{N}+\sqrt{2}\right)}+(\frac{1}{\mathcal{N}}+\sqrt{2})\frac{f_2}{f_1} + \mathcal{O}(\delta^1)
\end{split}
\end{equation}
By setting the number of E-fold into around 60, we can recalculate the initial  value of this inflation model $\theta_{Begin}$. Notice that the final term depends on the dimensionless fraction $\frac{f_1}{f_2}$.
\section{Minimal path in two scalar fields potential}
Considering now the full potential (\ref{pot}) and taking the case of $\alpha$ being very small, in this case, we have two scalar fields with two different characteristics. One of them ($\theta_2$) is weakly coupled, because we assume that $\alpha$ is very small. In contrast, $\mathcal{N}$ is always in order of $1$, which means that $\theta_1$ is a strongly couple scalar field. The following picture therefore appears reasonable. The scalar field $\theta_2$ satisfies a slow roll field equation, since $ \frac{\partial V_\textbf{eff}}{\partial\theta_2} \sim \alpha $, and therefore very small. Since the other field $\theta_2$ is rather strongly coupled, as $\theta_2$ field slow rolls, the $\theta_1$ can oscillate around the minimum for $\theta_1 \approx \textbf{const}$. Or even simpler we take the approximation that as $\theta_2$ slow rolls, and $\theta_1$ field sits at the minimum of the potential, for any given value of the slowly rolling $\theta_2$ field:
\begin{equation}
\frac{\partial V}{\partial\theta_1} \approx 0
\end{equation}
with the solution for $\theta_2$:
\begin{equation}
\theta_2  = \frac{\sqrt{3} \mathcal{N}}{\alpha} \ln \left[ \frac{3 \mathcal{N}^2 e^{-2 \mathcal{N} \theta_1} \left(f_1 e^{\mathcal{N} \theta_1}-2 f_2\right)}{M} \right] - \frac{\theta_1}{\sqrt{3}\alpha}
\end{equation}
For another example where a scalar field is taken  at any time in the
evolution of the universe to be sitting instantaneously at the minimum of
the effective potential see \cite{Fardon:2003eh}.

In order to calculate the number of E-folds in that approximation, lets calculate the slow roll parameters, with respect to the field $\theta_2$, where $\frac{\partial V}{\partial\theta_1} \approx 0$ is taken to be on the solution:
\begin{equation}
\epsilon \approx \frac{1}{2}\left(\frac{V_{,\theta_2}}{V}\right)^2_{\frac{\partial V }{\partial \theta_1}\approx 0}, \quad \eta \approx \frac{V_{,\theta_2,\theta_2}}{V}_{\frac{\partial V }{\partial \theta_1}\approx 0}
\end{equation}
and therefore:
\begin{equation}
\epsilon \approx \frac{\alpha ^2}{2}  \left(\frac{f_1 e^{\frac{\theta_1}{\sqrt{3}}}-2 f_2}{f_2-\lambda  e^{\frac{2 \theta_1}{\sqrt{3}}}}\right)^2+O\left(\alpha^4\right)
\end{equation}
\begin{equation}
\eta \approx \alpha ^2  \frac{M}{-f_1+f_2 e^{-\frac{\theta_1}{\sqrt{3}}}+M+\lambda  e^{\frac{\theta_1}{\sqrt{3}}}}+O\left(\alpha^3\right)
\end{equation}
The slow roll parameters are very small in a wide range of the scalar field, since the assumption of $\alpha$ is very small.

In order to estimate the end of inflation, we approximate $\epsilon$ to one, which gives the solution:
\begin{equation}
\begin{split}
\theta_{2}^\textbf{final} \approx  \frac{\sqrt{3}}{2} \ln \left(\frac{f_2}{\lambda}\right) + (1-\frac{f_1}{2\sqrt{f_2 \lambda}}) \sqrt{\frac{3}{2}}\alpha
\end{split}
\end{equation}
The number of E-fold is calculated from the integral:
\begin{equation}
\begin{split}
\textbf{NumE-folds} \approx \int_{\theta_\textbf{final}}^{\theta_{\textbf{begin}}} \frac{d\theta_2}{\sqrt{2\epsilon}}|_{\frac{\partial V}{\partial\theta_1} \approx 0}
\end{split}
\end{equation} 
Under the assumption of $\alpha$ being very small we get that the solution for the number of E-folds for that model is:
\begin{equation}
\textbf{NumE-folds}  \approx \frac{\mathcal{I}}{\sqrt{2}\alpha^3} + \mathcal{O}(\frac{1}{\alpha_2})
\end{equation}
where:
\begin{equation}
\begin{split}
\mathcal{I} =|\frac{2}{\sqrt{3}} \theta _{\text{Begin}} - \sqrt{3} (\mathcal{I}_1 - \mathcal{I}_2) - \mathcal{I}_3|
\end{split}
\end{equation}
\begin{equation*}
\mathcal{I}_{1} = 2 \ln \left[\frac{e^{-\frac{2 \text{Begin}}{\sqrt{3}}}}{M} \left(f_1 e^{\frac{\text{Begin}}{\sqrt{3}}}-2 f_2\right)\right]
\end{equation*}
\begin{equation*}
\mathcal{I}_{2} = \frac{(3f_1  e^{\frac{\text{Begin}}{\sqrt{3}}}-7 f_2+\lambda  e^{\frac{2\text{Begin}}{\sqrt{3}}}) (-f_1 e^{\frac{\text{Begin}}{\sqrt{3}}}+f_2+\lambda  e^{\frac{2\text{Begin}}{\sqrt{3}}})}{\left(f_1 e^{\frac{\text{Begin}}{\sqrt{3}}}-2 f_2\right)^2}
\end{equation*}
\begin{equation*}
\mathcal{I}_3 = \frac{\lambda }{f_2 M^2}\left(f_1 \sqrt{\frac{f_2}{\lambda }}-2 f_2\right)^2+3
\end{equation*}
From those terms we can estimate the initial value of the scalar field, in we fix the number of E-folds to be around 60.

\section{Comparison with the metric-affine formalism solution}
In the metric affine formalism the $\theta_1$ field does not exist, because the $\theta_1$ field equation is a constraint equation, which allows us to solve $\chi$ as a function of $U$ and $V$. This case for $\lambda =0$ was studied is \cite{Guendelman:1999qt}, and for opposite value of $f_2$. So the case $M=0$ is trivial, because it just corresponded to a constant potential case. On the other hand for $M \neq 0$ an inflaton field is obtained, but no additional field is arises.

\section{Conclusions}
In this paper we studied the formulation of scale invariant Two Measures Theory with a dilaton field. We similarly looking action to that already studied in the first order formulation, augmented by the addition of a $\Phi^2$ term in the action to generate an independent contribution to the cosmological constant. Although the actions are similar, the degrees of freedom are different, since in the second order formulation the ratio of the measure  $\Phi$ to the $\sqrt{-g}$ becomes a new degree of freedom, a new scalar field , absent in the first order formulation, where such a field satisfies a constraint equation instead of a dynamical equation, so that it can be eliminated in terms of other fields of the theory (like the dilaton field, etc.). An effective potential for two scalar fields is obtained, only one term in the potential, proportional to a scale symmetry breaking integration constant $M$ involves the dilaton field.

The theory in the absence of spontaneous symmetry breaking ($M=0$) contains a potential that depends on only one field, while the one represents a flat direction. We first study the possibility of inflation in the scale invariant limit and find that it is possible, no obstacles are found to obtain $60$ e-folds.

A second scenario, for $M\neq0$ involves the two scalar fields, choosing the scale coupling constant $\alpha$ to be small, we obtain that one of the fields can slow roll, while the other can either rapidly oscillate around the local minimum valid while taking the previous field constant, or in a more simple scenario, adjust at any given time to be at the minimum (for a given value of the other field), in this simple scenario also there will be no problems to achieve enough e-foldings for small enough $\alpha$. In the future, the primordial power spectrum for both models should be investigated as well as data fitting and constraints the parameters in the models. 
\acknowledgments
This article is supported by COST Action CA15117 "Cosmology and Astrophysics Network for Theoretical Advances and Training Action" (CANTATA) of the COST (European Cooperation in Science and Technology). We would like thank to Hendrik Edelmann for improving the plot design.

\end{document}